# Multiscale Topological Properties Of Functional Brain Networks During Motor Imagery After Stroke

Fabrizio De Vico Fallani[a,b,d], Floriana Pichiorri[a], Giovanni Morone[a], Marco Molinari[c] Fabio Babiloni[b], Febo Cincotti[a] and Donatella Mattia[a]

[a] *Neuroelectrical Imaging and BCI laboratory, IRCCS Fondazione Santa Lucia, Rome, Italy;*

[b] *Department of Physiology and Pharmacology, University Sapienza, Rome, Italy;*

[c] *Experimental Neurorehabilitation laboratory, IRCCS Fondazione Santa Lucia, Rome, Italy;*

[d] *Brain and Spine Institute (CRICM), UPMC/Inserm UMR_S975/CNRS UMR7225, Paris, France*

Corresponding author
Fabrizio De Vico Fallani, PhD, BME.
Email: fabrizio.devicofallani@gmail.com

## Abstract

In recent years, network analyses have been used to evaluate brain reorganization following stroke. However, many studies have often focused on single topological scales, leading to an incomplete model of how focal brain lesions affect multiple network properties simultaneously and how changes on smaller scales influence those on larger scales. In an *EEG*-based experiment on the performance of hand motor imagery (MI) in 20 patients with unilateral stroke, we observed that the anatomic lesion affects the functional brain network on multiple levels. In the beta (13-30 Hz) frequency band, the MI of the affected hand (Ahand) elicited a significantly lower smallworldness and local efficiency ($E_{loc}$) versus the unaffected hand (Uhand). Notably, the abnormal reduction in $E_{loc}$ significantly depended on the increase in interhemispheric connectivity, which was in turn determined primarily by the rise in regional connectivity in the parieto-occipital sites of the affected hemisphere. Further, in contrast to the Uhand MI, in which significantly high connectivity was observed for the contralateral sensorimotor regions of the unaffected hemisphere, the regions that increased in connection during the Ahand MI lay in the frontal and parietal regions of the contralaterally affected hemisphere. Finally, the overall sensorimotor function of our patients, as measured by Fugl-Meyer Assessment (FMA) index, was significantly predicted by the connectivity of their affected hemisphere. These results increase our understanding of stroke-induced alterations in functional brain networks.

## Abbreviations

*MRI* = Magnetic Resonance Imaging; *PET* = Positron Emission Tomography; *MEG* = MagnetoEncephaloGraphy; *EEG* = ElectroEncephaloGraphy; *DTI* = Diffusion Tensor Imaging; *Ahand* = Affected hand; *Uhand* = Unaffected hand; *Ahemi* = Affected/Ipsilesional hemisphere; *Uhemi* = Unaffected/Contralesional hemisphere; *MI* = Motor imagery; *FC* = Functional connectivity

# 1 Introduction

Most brain functions result from the organization of several neuronal assemblies in a complex and dynamic system (Varela et al., 2001). The term "organization" can be defined as the coherent interdependence of various parts that constitute the whole. Functional connectivity (*FC*) approaches have been introduced to operationally describe the temporal dependence across spatially remote neurophysiological processors (Friston, 1994); such approaches are effective tools for assessing the organization of the brain, based on the activity of multiple cerebral regions.

Over the past decade, graph theory has been introduced as a mathematical approach to characterize the complexity of anatomic and functional brain networks (Bullmore and Sporns, 2009). In functional neuroimaging, a graph is an abstract representation of a pattern of connectivity, in which nodes represent various areas of the brain and links correspond to significant interactions between the activities of regions of the brain. Many groups have exploited graph-based approaches to examine the changes in functional (data-driven) and effective (model-based) connectivity in several brain disorders (He and Evans, 2010). In this regard, many computational studies have focused on understanding how the brain reorganizes its functional structure after stroke from a network-based perspective.

Graph theory approaches have allowed the effect of stroke on the organization of the brain to be studied from brain signals that are recorded during resting states and task-related connectivity through various noninvasive techniques, such as functional *MRI (fMRI)* (Nomura et al., 2010; Wang et al., 2010), *EEG* (De Vico Fallani et al., 2009), *MEG* (Westlake et al., 2012), and *DTI* (Crofts et al., 2011). Although the extent to which the application of such approaches impact the study of stroke-related disturbances in cortical connectivity is unknown, they have been reviewed comprehensively, based on recent meta-analyses (Grefkes and Fink, 2011; Westlake and Nagarajan, 2011).

These reviews have highlighted that stroke lesions can effect *i)* critical deviation from optimal "small-world" network topologies that support processing of segregated and integrated information (Bassett and Bullmore, 2006), *ii)* altered interhemispheric connectivity, *iii)* and abnormal region centrality in the ipsilesional and contralesional hemispheres, possibly due to compensatory mechanisms. Although this evidence suggests that stroke modulates several topological attributes of the functional brain network, ranging from small (eg, single-node connectivity) to large scales (eg, connectivity of the entire system), a unifying framework that simultaneously describes the changes in network properties on different scales (Alstott et al., 2009) and their relationships (Vázquez et al., 2004) has not been established and is rarely and poorly applied in analyses of functional brain networks.

In this study, we applied a multilevel graph analysis method of functional brain networks that was built from *EEG* signals and designed to examine multiple topological scales simultaneously. Specifically, based on the peculiarity of functional brain networks that are to be embedded in a physical space that is coincident with the anatomic substrate (Honey et al., 2007; Doron et al., 2012), we aimed to characterize the *FC* patterns on several scales: *i)* the entire brain (large scale), *ii)* the 2 hemispheres (intermediate scale), and *iii)* each node in the 2 hemispheres (small scale). This framework was adopted to describe the possible brain connectivity disturbances in stroke.

Specifically, the functional brain network was studied under a task-specific condition, represented by mental simulation of hand movements, also called motor imagery (*MI*). *MI* can be defined as a dynamic state during which the representation of a specific motor action is rehearsed internally without any overt motor output and is governed by the principles of central and peripheral motor control (Decety, 1996). Thus, the practice of action mental imagery by triggering neural activations of relevant brain motor areas is an alternative approach for examining the motor system, even in the absence of movement execution (Page et al., 2007; Sharma et al., 2006).

Based on these considerations, we used the proposed graph approach to study the functional brain networks in stroke patients with unilateral cortico-subcortical damage of the sensorimotor system that caused various degrees of motor impairment in the respective contralateral side (ie, hemiplegia or hemiparesis). The patients performed *MI* with their affected (*Ahand*) and unaffected (*Uhand*) hands, the latter of which was used as the reference condition (Jang et al., 2003; Johansen-Berg et al., 2002) to be contrasted to the *Ahand* MI representing the target condition under investigation. *FC* was estimated from scalp *EEG* signals, which have high temporal resolution and carry frequency-specific information on motor task-related neural activity (Babiloni et al., 1999; Gerloff et al., 1998).

We hypothesized that our experimental design would allow us to:

- assess the impact of unilateral stroke lesions on multiple brain network properties, estimated during the mental rehearsal of movements, and identify possible dependencies between the network changes on various topological scales; and
- examine the presence of reliable network-based neuromarkers that correlate with poststroke functional motor status, as measured using motor functional scales.

## 2 Materials and Methods

Between 2011 and 2012, we recruited 20 patients (mean age, *55.5* years; 11 females) who were affected by a first-ever unilateral stroke in the subacute phase (time since event, *8.4±2.8* weeks) on admission for poststroke rehabilitation treatment at Fondazione Santa Lucia (Rome). All patients had suffered unilateral supratentorial (cortico/subcortical) stroke (left hemisphere *11*) that was confirmed by structural MRI and resulted in various degrees of motor impairment on the side of the body that was contralateral to the stroke lesion (for patient details, see **Tab. 1**). Exclusion criteria were: the pharmacological treatment with drugs affecting the patient's vigilance and/or the *EEG* background activity; Mini-Mental State Examination score < 24 (Tombaugh, 2005) and severe cognitive disorders (such as severe hemispatial neglect and language disorders) as evaluated by a neuropsychologist; the presence of other chronic disabling pathologies; orthopedic injuries that could impair reaching or grasping; spasticity of the shoulder, elbow, or finger flexors and extensors that exceeded 3 on the modified Ashworth Scale;

The clinical and functional assessment of all patients comprised the following scales: *i)* the European Stroke Scale (Hantson et al., 1994); *ii)* the Medical Research Council scale for muscle strength (Compston, 2010) to assess residual strength in the upper limbs; and *iii)* the upper limb section of the Fugl-Meyer Assessment (Fugl Meyer et al., 1975) to assess functional motor recovery after stroke. Detailed scale scores relative to the clinical and functional assessment are reported in **Tab. 1**. All measurements were made by an expert physician less than 1 week before the *EEG* data acquisition. All patients gave written informed consent for participation in the study, which was approved by ethical committee of the Fondazione Santa Lucia.

### 2.1 EEG Recordings and Motor Tasks

All patients had *EEGs* recorded within 1 week after hospitalization. Patients were comfortably seated in a dimly lit room, with their upper limbs resting on a cushion, and instructed by a visual cue to perform a kinesthetic type of *MI* of their hand grasping (Jeannerod, 1994).

**Table 1 - Demographic, clinical, and functional characteristics of the stroke patients**

| Patient | Age | Sex | Hand plegia/week | Lesion side | Lesion Type | MRC | FMA | ESS |
|---|---|---|---|---|---|---|---|---|
| #1 | 43 | F | 12 | L | left fronto-parietal,basal ganglia,amigdala ischemia | 53 | 18 | 56 |
| #2 | 51 | F | 8 | L | left fronto-temporo-parietal ischemia | 45 | - | 57 |
| #3 | 45 | F | 12 | R | right fronto- temporo-parietal ischemia | 51 | - | 63 |
| #4 | 53 | F | 12 | R | right temporal and basal ganglia haemorragia | 54 | - | 65 |
| #5 | 41 | F | 8 | L | Left-fronto-parietal-rolandic convexity ischemia | 77 | - | 89 |
| #6 | 47 | F | 8 | R | right fronto-temporo-parietal ischemia | 44 | - | 66 |
| #7 | 66 | F | 12 | R | right fronto-temporal-parietal ischemia | 50 | 17 | 63 |
| #8 | 41 | M | 12 | R | right fronto-temporo-parietal ischemia | 49 | - | 65 |
| #9 | 64 | M | 6 | R | right nucleo-capsular ischemia | 46 | 10 | 56 |
| #10 | 70 | M | 5 | R | right mca thrombosis with ischemia | 78 | 60 | 96 |
| #11 | 54 | M | 7 | R | right nucleo-capsular, temporal lobe ischemia | 76 | 49 | 90 |
| #12 | 70 | M | 5 | R | left nucleo-capsular, temporal subcortical | 46 | 8 | 47 |
| #13 | 57 | M | 6 | R | right emi pons ischemia - | 72 | 44 | 78 |
| #14 | 62 | M | 12 | L | left fronto-temporo-parietal ischemia | 72 | 54 | 89 |
| #15 | 64 | M | 6 | L | left fronto-mesial, insular ischemia | 70 | 37 | 82 |
| #16 | 71 | F | 6 | L | left emipons ischemia - | 72 | 61 | 76 |
| #17 | 75 | F | 10 | L | left-cortical-subcortical-fronto-insular,prerolandic ischemia | 72 | 44 | 75 |
| #18 | 58 | M | 10 | L | semioval center and corona radiata | 60 | 21 | 66 |
| #19 | 34 | F | 8 | L | fronto-temporoinsular cortical-subcortical | 43 | 9 | 62 |
| #20 | 44 | F | 4 | L | nucleobasal-insular left - | 41 | 5 | 47 |
| Mean | 55,5 | | 8,4 | | | 58,6 | 31,2 | 69,4 |
| St.Dev. | 11,9 | | 2,8 | | | 13,4 | 20,7 | 14,4 |

*F=Female, M=Male, R=Right, L=Left. ESS = European Stroke Scale: the scale ranges from 0 (maximally affected person) to 100 (normal). MRC = Medical Research Council scale for muscle strength, upper limbs section: the scale ranges from 0 (no movement) to 5 (complete movement against full resistance) for each segment explored (8 segments per side in the upper limb). FMA = Fugl-Meyer Assessment: scores range from 0 (maximally affected) to 66 (normal); FMA was performed in 14 of the 20 recruited patients.*

In order to ensure the correct understanding of the *MI* task by the patients, several trials of actual execution of the same sustained grasping with the unaffected hand were performed before the recording session (visual cue and timing as the *EEG* experimental condition). Afterwards, in the *EEG* experiment, patients were instructed to rehearse "the feeling of movements" acquired during the previous *MI* task practice. Similar pre-*EEG* recording session practice was allowed with the affected hand by attempting grasping movements. The recording session comprised 2 runs in which the *MI* of the hand grasping relative to the unaffected (*Uhand*) and affected (*Ahand*) hand was sustained for 4 sec. Each run consisted of *30* trials (8 s each), divided equally between randomly presented *baseline* and *task* trials. The visual cue was presented using dedicated software, ie,BCI2000 (Schalk et al., 2004), that was synchronized with the *EEG* amplifiers.

As illustrated in **Fig. 1,** the visual cue was a small red ball that moved at constant speed along the central vertical line of a screen from bottom to top for *8 s* (trial duration). In the *task* trials (panel **A**), the lower half of the screen was black and the upper half was green. Patients were instructed to be prepared to begin the hand *MI* as soon as the red ball entered the green area (*4 s*) and maintain the task until the ball reached the edge of the screen (*4 s*). In the *baseline* trials (panel **B**), the screen was black, and patients simply relaxed throughout the trial duration (*8 s*).

*EEG* signals were collected from *61* scalp sites that were assembled on an electrode cap per a montage that was modified as an extension of the international 10-20 system. The electroculogram (*EOG*) was simultaneously recorded to allow the subsequent rejection of ocular artifacts. *EEG* data were continuously acquired on a commercial system (Brainproduct GmbH, Munich, Germany) with *200* Hz frequency sampling that was referenced to the linked-ear signal. The data were then band-pass filtered in the *1-45 Hz* range and depurated from ocular artifacts using the Independent Component Analysis tool (ICA) and commercial software (Vision Analyzer software; Brainproduct GmbH, Munich, Germany).

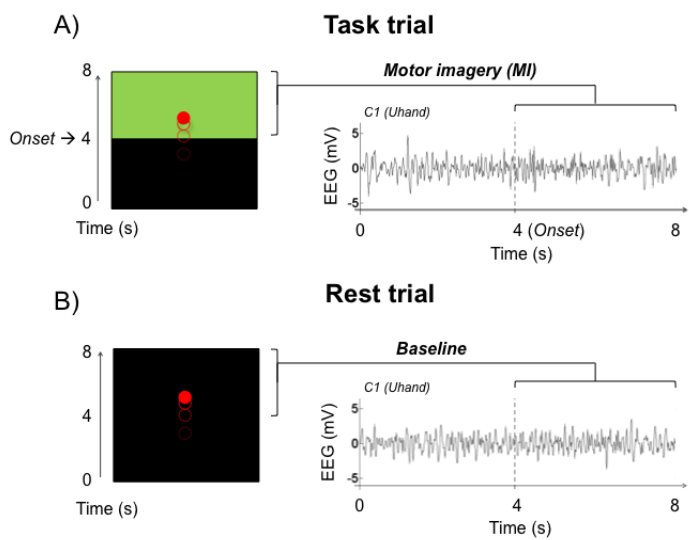

**Figure 1 - Schematic illustration of a representative experimental session.**
*The visual cue is shown on the left side of panel A (Task trial) and B (Rest trial). The right side of panel A and B shows the raw EEG signal recorded from a patient performing the motor imagery of the unaffected hand (Uhand). Traces were obtained from C1 electrode position and were relative to a Task and a Rest trial. In both cases, the trial duration was 8 s and the period of interest for functional connectivity estimation ranged from the 4th to the 8th second.*

To ensure that the *MI* task was performed without any concomitant voluntary muscular contraction, the electromyographic (*EMG*) activity that was recorded in the left and right opponens pollicis was monitored throughout the experimental session using disposable surface electrodes that were placed in a bipolar belly-tendon configuration. The *EMG* signals were available to the experimenter to encourage the patients to relax their muscles and avoid movements during the task trials. The preprocessed *EEG* signals were then segmented considering the last *4* s of each *task* and *baseline* trial as the period of interest, as shown in **Fig. 1**. The segmented traces were inspected visually to reject any *EEG* segment that had residual muscular or other pronounced artifacts. In the offline analysis, we flipped the functional (*EEG* time series) and anatomic (scalp electrode positions) data of patients with left-sided lesions along the mid-sagittal plane to perform a group analysis with all 20 patients.

**2.2 Functional Connectivity Estimation**

Brain *FC* was calculated for the segmented periods of interest (*task* and *baseline*) using imaginary coherence (Nolte et al., 2004), which is a robust estimate of synchronization between 2 time series in the frequency domain. This method yields an *FC* that is unaffected by the volume condition noise due to the anisotropic conductivity of the skull, which blurs the original signals that are generated by the cortical surface. Imaginary coherence gives weighted values between *0* and *1* for each frequency—ie, higher imaginary coherence values in a frequency reflect greater synchronization between the *EEG* oscillations at that frequency. The original values of imaginary coherence were then Z-transformed to ensure that they approximated a normal distribution (Nolte et al., 2004).

To study the level of synchronization in specific physiological frequency contents, the Z-transformed imaginary coherence values were averaged within specific ranges, yielding a single mean value that characterized various EEG bands of interest—namely *Theta* (*4-8 Hz*), *Alpha* (*8-13 Hz*), *Beta* (*13-30 Hz*), and *Gamma* (*30-40 Hz*). For each patient, run, and frequency band, the connectivity patterns of the *MI* task segments were contrasted statistically with those of the baseline segments. The statistical comparison was performed over all possible electrode pairs by paired nonparametric Wilcoxon signed rank test, denoted here as *W-test*.

The functional brain networks that characterized the *MI* were obtained by maintaining the coherences whose values in the task differed significantly from those at baseline. Similar procedures have been proposed and used in previous studies, in which the functional brain network was the result of a statistical comparison between conditions (Ginestet and Simmons, 2011; Zalesky et al., 2010) or populations (De Vico Fallani et al., 2010). For an analysis of resting states, in which no contrasting procedure is available, the general procedure consisted of repeating the network analysis for a series of increasing threshold values (Rubinov and Sporns, 2010). The significance threshold was set to *p=0.05* and adjusted for multiple comparisons by rough false discovery rate (RFDR) correction, based on the number of node pairs for which the *W-tests* were computed (*p* adjusted to *0.025*). Recently, the RFDR criterion has been applied in several neuroimaging studies (Supekar et al., 2008; Wolf et al., 2011).

**2.3 Network Analysis**

The estimated *FC* patterns were characterized using network metrics that have been derived from graph theory (Costa et al., 2011). In our graphs, nodes represent scalp electrodes (*N=61*) and links represent coherence values between pairs of *EEG* signals. To eliminate any topological bias due to disparate connection densities between brain networks (van Wijk et al., 2010), we transformed the significant coherence values (from the comparison with baseline segments) into binary values and decreased the links in each brain network to the minimum number that was common to all patients, conditions, and frequency bands.

Thus, the original sparse weighted graphs were converted into unweighted graphs by retaining only the *185* most significant links (*L=185*) and transforming them into binary values—ie, *0* = no link and *1* = presence of significant link. Although neglecting the weight of the links (ie, the coherence value) can be considered a reduction of the available information, we noticed that the interpretation and use of link weights in brain network analyses remain an controversial issue, particularly due to the undefined relationship with the concept of physical distance in graphs (Rubinov and Sporns, 2010).

A multilevel topological analysis was eventually applied to the estimated brain networks.

*2.3.1 Large Scale*

Two indices were considered to study the more coarsely grained features of the brain networks - **global** and **local efficiency** ($E_{glo}$ and $E_{loc}$, respectively) - which have been used extensively to characterize the global properties of functional brain networks (Bullmore and Sporns, 2012). $E_{glo}$ and $E_{loc}$ reflect the same properties of the inverse of the average shortest path *L* and clustering index *C*, which were introduced by Watts and Strogatz to characterize the small-world property of networks (see **Appendix A** for details**).** A simple measure of efficiency-based smallworldness, $SW$, can be calculated as $SW = \frac{E_{loc}/E_{loc}r}{E_{glo}/E_{glo}r}$ where $E_{loc}r$ and $E_{glo}r$ are the mean efficiency values from equivalent random graphs (Downes et al., 2012). Whenever *SW>1*, a network is considered to exhibit small-world properties.

*2.3.2 Intermediate Scale*

The intermediate-scale level was addressed by examining the properties of 2 predetermined sets of nodes, $S_{Ahemi}$ and $S_{Uhemi}$, which correspond to the scalp sensors of *Ahemi* and *Uhemi*, respectively (see **Appendix A** for details).

**Interdensity $K_{inter}$** is defined as the actual number of links that run between the 2 sets over all possible edges between them. By definition, interdensity ranges from *0* to *1*, wherein higher $K_{inter}$ values reflect a greater number of interhemispheric links. We considered interdensity to be a particular instance of *cut size*, a graph index that is used frequently to determine the optimal partition, consisting of separate clusters (see Supplementary **Text,** Section **S1.1**).

**Intradensity $K_{intra}$** is defined as the ratio between the actual number of links in a set and the total number of possible links in the same set. By definition, intradensity ranges from *0* to *1*; higher $K_{intra}(S)$ values indicate greater connection between the nodes in set *S*.

*2.3.3 Small Scale*

On the small-scale level, we considered graph indices that extracted the finer-grained properties of the *EEG* network. Two indices were defined to measure the centrality of nodes with respect to the connectivity between and within $S_{Ahemi}$ and $S_{Uhemi}$ (see **Appendix A** for details). These sets were symmetric and thus had the same number of nodes - $N_{SAhemi}=N_{SUhemi}=N_S=26$ electrodes.

**Interdegree $D_{inter}$** was computed as the total number of links of a node in a set to those of the other set. By definition, $D_{inter}$ ranges from *0* to $N_S$. A node with high $D_{inter}$ is considered central, because its removal would reduce overall interhemispheric connectivity.

**Intradegree $D_{intra}$** was computed as the total number of connections of a node to other vertices in the same group. By definition, it ranges from *0* to $N_S-1$. A node with high $D_{intra}$ is considered central, because its removal would decrease overall intrahemispheric connectivity.

*2.3.4 Normalization by random graphs*

To handle all normalized network indices, we referred to completely random connectivity patterns, in which links were arranged randomly. Notably, *1000* random graphs were generated by maintaining the same number of nodes and links of the original brain networks. In each instance, links were shuffled randomly without preserving the distribution of node degrees (Sporns and Zwi, 2004). Ultimately, all graph measures that were computed from various brain networks were divided by the respective mean values from the random graphs. When this ratio is lower than 1, the generic brain network property is lower than random graphs; when the ratio exceeds 1, it is higher than random graphs.

*2.3.5 Statistical comparison between conditions*

We use paired nonparametric Wilcoxon signed rank test (*W-test*)*,* with a statistical threshold of *0.05,* to analyze the differences between brain network indices that were computed for *Ahand* and *Uhand* conditions (this latter was considered as our reference condition). For small-scale topologies, in which local measures were computed for each node, we corrected for multiple comparisons. Specifically, a nominal significance of *p=0.05* was defined and adjusted for multiple comparisons by RFDR correction, based on the number of nodes for which the *W-tests* were computed (*p* adjusted to *0.0255*). RFDR is a less restrictive procedure for multiple comparisons with greater power than family-wise error rate (FWER) control at the cost of increasing the likelihood of obtaining type I errors (Zar, 1999).

**2.4 Interscale dependence between brain network properties**

In this study, we also examined the dependence of the brain network properties at larger topological scales on those of smaller scales—ie, small -> intermediate, intermediate -> large. The linear regression coefficient was computed between the respective network values (independent variables = larger-scale values, dependent variables = smaller-scale values) from all patients under both conditions (*Ahand* and *Uhand*) and for each frequency band. Notably, only network attributes that had already reported significant differences between conditions (section **2.3.5**) were considered for regression analysis. Evaluation of the regression coefficients determined whether and how changes in finer-grained network properties (smaller scale) influence or "possibly cause" (Sokal and Rohlf, 1994) changes in coarser-grained properties (larger scale).

*2.4.1 Intrinsic relationships between graph indices*

When disparate graph indices are estimated on the same network, they could have a high degree of correlation as a simple consequence of their intrinsic topological definitions—eg, local efficiency and clustering coefficient (Latora and Marchiori, 2001). In this study, we limited such phenomena by considering graph indices that characterized different scales of topology. Nevertheless, these basic relationships should be determined to interpret the interdependence between changes in graph indices fairly. In general, node degrees (intra/inter) are intuitively related to the connection density (intra/inter)—ie, a set of nodes with a higher degree indicates greater connection density.

There is a less defined relationship between efficiency values and connection densities. In particular, global ($E_{glo}$) and local efficiency ($E_{loc}$), which reflect integration and segregation tendencies between groups of nodes, respectively, are intended to be related to the connectivity between hemispheres—ie, $K_{inter}$. To examine this issue, we implemented a simulation model that characterized the dynamics of efficiency-based values (*SW*, $E_{glo}$, $E_{loc}$) with regard to increasing interhemispheric connectivity ($K_{inter}$).

The model generated a sequence of networks that had the same size of the brain networks that were considered here—ie, *N = 61* and *L=185*. Starting from a network configuration in which all *185* links were arranged randomly in 1 hemisphere ($N_S$=26 nodes), the model reassigned an increasing number of links between the hemispheres randomly until it reached a configuration that had only interhemispheric links. The choice of such model characteristics was suggested by recent evidence of the effects of stroke on interhemispheric connectivity (Grefkes and Fink, 2011; Westlake and Nagarajan, 2011).

To this end, we hypothesized that the estimated brain networks should lie within a range that is delimited by perfect hemisphere lateralization with respect to sensorimotor control of the contralateral hand (McFarland et al., 2000; Volkmann et al., 1998) and abnormal and complete interhemispheric connectivity. Briefly, an increasing number of links *l=1,2,...,185* was shuffled randomly in the simulation model. Because we did not know the optimal proportion of the *l* links to be rearranged between the hemispheres a priori, we introduced a parameter, $p_{inter}$, to vary the ratio.

Thus, the simulated network configuration exhibited interhemispheric links that were proportional to $p_{inter}$ and intrahemispheric links that were proportional to *1-$p_{inter}$*. When $p_{inter}$=0, we invoked the inferior limit condition in which the *l* links were reassigned in only 1 hemisphere. When $p_{inter}$=1, the *l* links were rearranged exclusively between hemispheres. To simplify this process, 6 equally spaced $p_{inter}$ values were selected: *0, 0.2, 0.4, 0.6, 0.8, and 1*. For each $p_{inter}$ value, *1000* random configurations were generated to obtain proper confidence intervals for the simulation model. More details can be found in **Appendix B**.

**2.5 Correlation with functional/clinical measures**

Statistical correlations were computed between functional/clinical measures in patients and values of the brain network indices that resulted significant after the *Ahand* versus *Uhand* contrast (section **2.3.5**). Specifically, for those network indices, we considered a delta index (*Δ*), calculated as the difference between the values for *Ahand* and *Uhand*. The *Δ* values were then used to determine the correlations with functional scale scores. The nonparametric Spearman correlation coefficient *R* was used to analyze the statistical dependences between the brain network indices and the scores on the functional scales. A *p*-level of *0.05* was the threshold for statistical significance. This statistical threshold was initially preferred to an adjusted level for multiple comparisons as we wanted to focus on few planned correlations between the Fugl-meyer assessment score (FMA), which is specific of the motor function of the patients, and the significant large-scale and intermediate-scale network indexes that resulted from *Ahand* versus *Uhand*. We are aware that this choice is arbitrary and though it reduces family-wise type II errors, it does not control for family-wise type I errors (Zar, 1999).

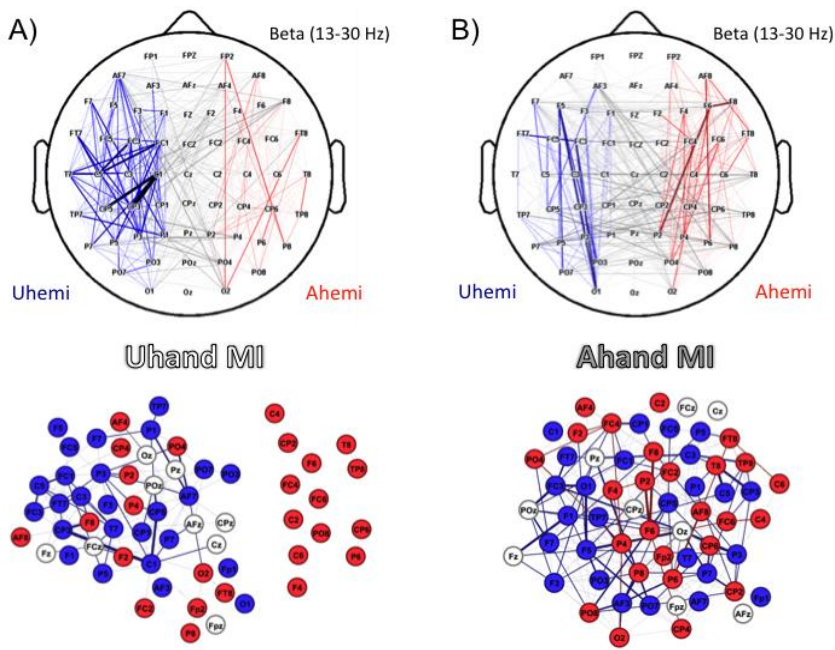

**Figure 2 - Grand average (n=20) of EEG network profiles in the Beta band during the MI of the unaffected Uhand and affected Ahand hand.**

*Top part: Grand average of the FC patterns relative to Uhand (panel A) and Ahand (panel B) condition. Blue and red lines denote the links within the unaffected (Uhemi) and the affected (Ahemi) hemisphere, respectively. Gray lines denote the inter-hemispheric links. The intensity of the color and the thickness of the lines vary as function of the number of patients exhibiting that significant link.*

*Bottom part: graph representation of the FC patterns relative to Uhand (panel A) and Ahand (panel B) condition. In this representation nodes are spatially repositioned through a force-based algorithm so that all the links are approximately of equal length with as few crossing edges as possible. Only links that were in common to more than 4 patients (20% of the sample) are illustrated here. Blue and red nodes indicate scalp electrodes placed over the undamaged (Uhemi) and damaged (Ahemi) hemisphere, respectively. The midline scalp electrodes (from Fpz to Oz) are illustrated as white nodes. Such graph representation highlights the existing partition in different clusters of the Brain networks.*

## 3 Results

**Fig. 2** shows the grand average (*n=20* patients) of the brain networks in the representative *Beta* band. Under the *Uhand* condition (panel A), the overall *FC* tended to converge on the contralateral (contralesional) hemisphere (*Uhemi*), primarily over the scalp sensorimotor area (electrode *C1*). In contrast, we did not observe specular behavior in *Ahand* (panel B), wherein the *FC* pattern had a similar distribution in the 2 hemispheres with a relatively high representation of frontoparietal and interhemispheric links (grey lines). This profile was also evident in the bottom section of the same figure, in which *FC* patterns are represented as graphs. Under the *Uhand* condition, the brain network (represented as a graph) appeared to segregate into 2 primary clusters of nodes, coincident with the 2 hemispheres. In contrast, a more intermingled structure of connectivity emerged in *Ahand*.

3.1 Multiscale *EEG* network properties

The analysis for the large-scale topological level was performed by comparing the *SW*, $E_{glo}$, and $E_{loc}$ indices under the *Ahand* versus *Uhand* conditions (see **2.3.5**). All estimated brain networks tended to have small-world properties—ie, $SW>1$ (**Fig. 3A**). In the *Beta* band, *SW* was significantly ($p=0.025$) lower in *Ahand* compared with *Uhand*. Similarly, in the *Beta* band, $E_{loc}$ under *Ahand* was significantly ($p=0.006$) lower than that under *Uhand* (**Fig. 3B**). $E_{glo}$ did not differ between conditions in any *EEG* frequency band (supplementary **Tab. S1**).

On the intermediate-scale level, we compared inter- and intrahemispheric *FC* separately between *Ahand* and *Uhand*. Interhemispheric connectivity, as measured by the interdensity $K_{inter}$, was significantly higher in *Ahand* for the *Beta* ($p=0.045$) band (**Fig. 3C**). Consistent with this observation, the tendency of the brain networks to form 2 separate clusters that were coincident with the 2 hemispheres was significantly lower (*Beta* $p<0.001$) in *Ahand* versus *Uhand* (see Supplementary **Text**, section **S2.1**).

In the analysis of intrahemispheric connectivity, the intradensity $K_{intra}$ of the unaffected hemisphere $K_{intra}(Uhemi)$ in the *Beta* band was significantly higher ($p=0.009$) in the contralateral *Uhand* with respect to *Ahand* (**Fig. 3D**). A nearly significant difference ($p\sim0.051$) was also observed for the $K_{intra}(Ahemi)$ values between the contralateral *Ahand* and *Uhand*. No other significant differences were noted for $K_{inter}$ or $K_{intra}$ in the other frequency bands (**Tab. 2**).

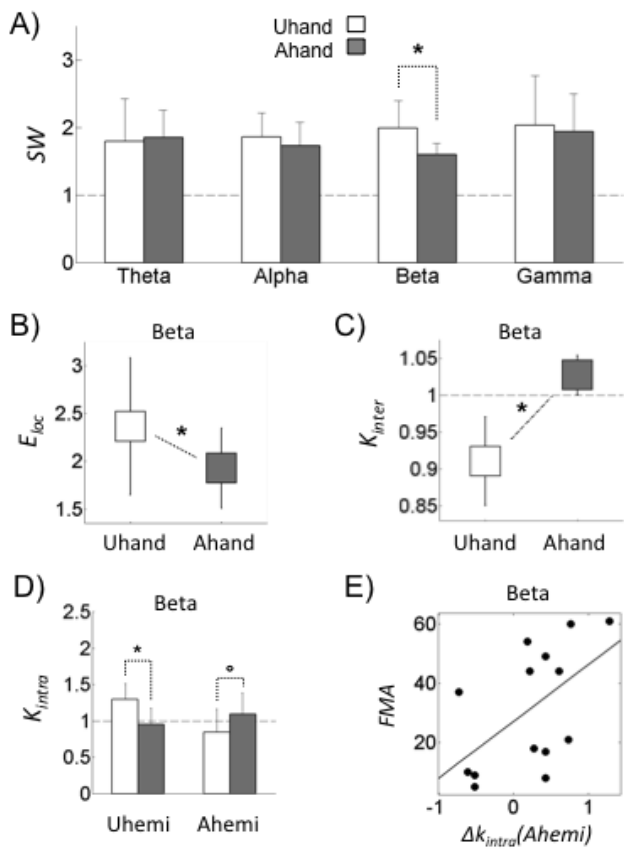

**Figure 3 - Large-scale and intermediate-scale brain network properties and correlation with functional/motor scale.**

*Panel A: Mean values of the smallworldness SW index estimated in the four EEG frequency bands for both the unaffected (Uhand, white bars) and affected hand (Ahand, grey bars) MI conditions. Vertical lines indicate the standard deviations. The asterisk denotes the significant ($p < 0.05$) differences between conditions. The same applies for vertical lines and asterisks in panel B, C and D. Panel B: Mean values of local-efficiency $E_{loc}$ obtained in the Beta band and relative to Uhand (white square) and Ahand (grey square) MI. Panel C: Mean values of inter-hemispheric density $K_{inter}$ obtained in the Beta band and relative to Uhand and Ahand MI. Panel D: Mean values intra-hemispheric density $K_{intra}$ relative to the unaffected Uhemi and affected Ahemi hemisphere in the Beta band obtained under both Uhand and Ahand MI condition. The white circle denotes a marginal significant ($p\sim0.05$) difference between conditions. Panel E: Scatter distribution of the stroke patient sample according to the behavioral sensorimotor Fugl-Meyer Assessment (FMA) and the variations (Δ) of intra-density (Kintra) of the affected hemisphere (Ahemi) in the Beta range of frequency. The $\Delta K_{intra}(Ahemi)$ scores are shown on the X-axis. Positive score values indicate that values are higher in Ahand than Uhand condition, the opposite being true for negative score values. The FMA scores are reported on the Y-axis. The lower is the FMA value the more severe is the degree of motor impairment and viceversa.*

At the smallest-scale level, the analysis was performed by comparing the interhemispheric and intrahemispheric *FC* of each scalp site separately between the 2 conditions. In the *Beta* band, the interdegree $D_{inter}$ of the interhemispheric local connectivity was significantly higher in *Ahand* versus *Uhand* for sites *PO8* (*p=0.01*) and *P8* (*p=0.02*) in the contralaterally affected hemisphere *Ahemi* (**Fig. 4A**).

In the same band, intradegree $D_{intra}$ of the intrahemispheric local connectivity was *i)* significantly higher in *Ahand* for sites *F2* (*p=0.017*)*, F4* (*p=0.027*)*,* and *P4* (*p=0.016*) of the contralateral *Ahemi* and in *Uhand* for sites *C1* (*p=0.019*)*, CP1* (*p=0.014*)*, CP3* (*p=0.006*)*,* and *C5* (*p=0.013*) of the contralateral *Uhemi* (**Fig. 4B).**

Few marginal significant differences were noted on the small-scale topological level in the other frequency bands (see Supplementary **Text**, section **S2.2**). We could not exclude that such differences could be affected by false positives as a consequence of the less restrictive procedure used to control multiple comparison (RFDR, see 2.2). However, these differences were considered to have minor relevance, because no concomitant significant differences were observed at the larger topological scales.

3.2 Influence of small-scale network properties on large-scale network properties

In the *Beta* band, in which significant differences between conditions have been reported on each topological scale, we observed a significant negative dependence of local efficiency $E_{loc}$ on interdensity $K_{inter}$ (*R=-0.342, p=0.030*), which, in turn, depended positively on the degree $D_{inter}$ of node *PO8* in the affected hemisphere *Ahemi* (*R=0.415, p=0.007*) (**Fig. 5**, lower central region). No significant influence of intradensity $K_{intra}$(*Uhemi*) or $K_{intra}$(*Ahemi*) on $E_{loc}$ was seen (**Fig. 5**, upper inner left and right regions). Yet*,* the intradensities of the 2 hemispheres exhibited a strong relationship with the degrees of specific nodes in the respective hemisphere*.* In the unaffected hemisphere, a significant positive dependence of $K_{intra}$(*Uhemi*) on $D_{intra}$ of node sites *C1*, *CP1,* and *CP3* appeared in the regression analysis *(C1 -> R=0.475, p=0.0003; CP1 -> R=0.426, p=0.006; CP3 -> R=0.569, p=0.0001)* (**Fig. 5**, upper outer left region)*.* Finally, in the affected hemisphere, there was a significant positive dependence of $K_{intra}$(*Ahemi*) on $D_{intra}$ of nodes *F2, F4 ,*and *P4* (*F2 -> R=0.455, p=0.001*; *F4 -> R=0.477, p=0.003; P4 -> R=0.547, p=0.0002)* (**Fig. 5**, upper outer right region). No other significant relationships were seen.

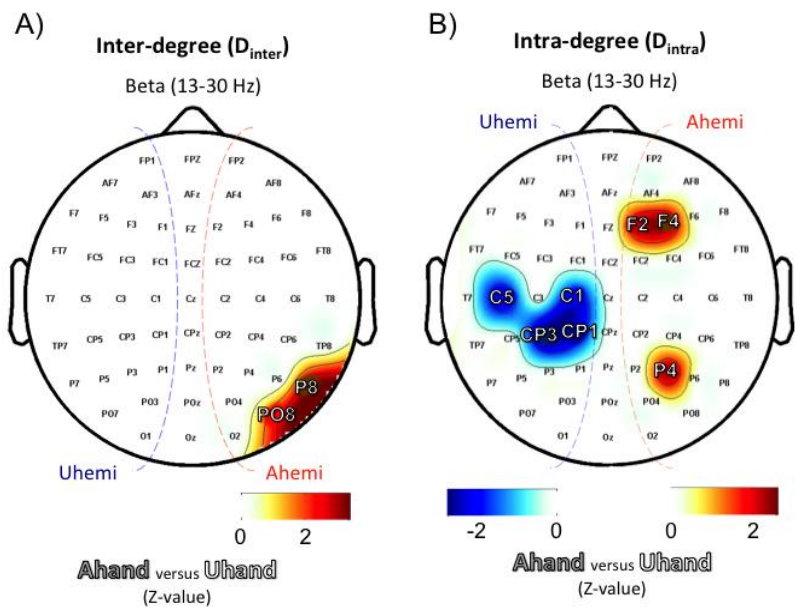

**Figure 4 - Scalp maps of statistical significant differences relative to small-scale network properties in the Beta band.** *The inter- degree $D_{inter}$ statistics are represented on panel A) and intra-degree $D_{intra}$ statistics are shown in panel B). Only the significant (p<0.05, RFDR corrected for multiple comparisons) Z-values resulting from the contrast (Wilcoxon-test) Ahand-versus-Uhand MI conditions are shown at each electrode position. Red colors stand for significantly higher index values in Ahand MI (Z>0) condition, while blue colors represent the reversal (Z<0). Shading colors are used for illustrative purposes.*

|  |  | $K_{intra}$ | | | | $K_{inter}$ | |
|---|---|---|---|---|---|---|---|
|  |  | *Uhemi* | | *Ahemi* | | - | - |
|  |  | *Uhand* | *Ahand* | *Uhand* | *Ahand* | *Uhand* | *Ahand* |
| **Theta** | Mean | 1,06 | 0,91 | 0,99 | 1,04 | 0,93 | 1,03 |
|  | SD | (0,20) | (0,13) | (0,14) | (0,25) | (0.2) | (0.21) |
| **Alpha** | Mean | 1,16 | 1,00 | 0,99 | 0,98 | 0.93 | 0,99 |
|  | SD | (0,43) | (0,16) | (0,16) | (0,29) | (0.19) | (0.25) |
| **Beta** | Mean | 1,29 * | 0,95 * | 0,84 ° | 1,09 ° | 0.91 * | 1,03 * |
|  | SD | (0,21) | (0,21) | (0,33) | (0,29) | (0.22) | (0.16) |
| **Gamma** | Mean | 0,87 | 1,02 | 1,12 | 1,02 | 0.98 | 1,05 |
|  | SD | (0,21) | (0,13) | (0,17) | (0,21) | (0.27) | (0.20) |

**Table 2 - Mean values of intermediate-scale brain network indexes for each condition and frequency band.**

*Standard deviations are reported within parentheses. Asterisks indicate significant differences (p<0.05) between the values of the Ahand and Uhand condition. White circles stand for marginal significantly (p~0.05) different values between Ahand and Uhand.*

*3.2.1 Model validation*

Whereas the dependence between node degrees and connection densities was intuitive (ie, higher node degrees contribute to higher connection density), the relationship between interdensity and local efficiency was apparently more complex and examined using the simulation model from Section **2.4.1**. Within the range of values for the model parameter $p_{inter}$ that controls the proportion of links that are rearranged randomly between hemispheres, we first reported that $p_{inter}=0.8$ reproduced the efficiency-based values and interdensities that corresponded to the actual values for the estimated brain networks (see interval of shuffled links between *75* and *90* in **Fig. S1** and **Tab. S1**). Further, in this optimal model configuration and specific interval of redistributed connectivity, the linear increase in interdensity (black line) caused a nearly linear decline in $E_{loc}$ (red line), whereas $E_{glo}$ values (blue line) remained stable (**Fig. 6**).

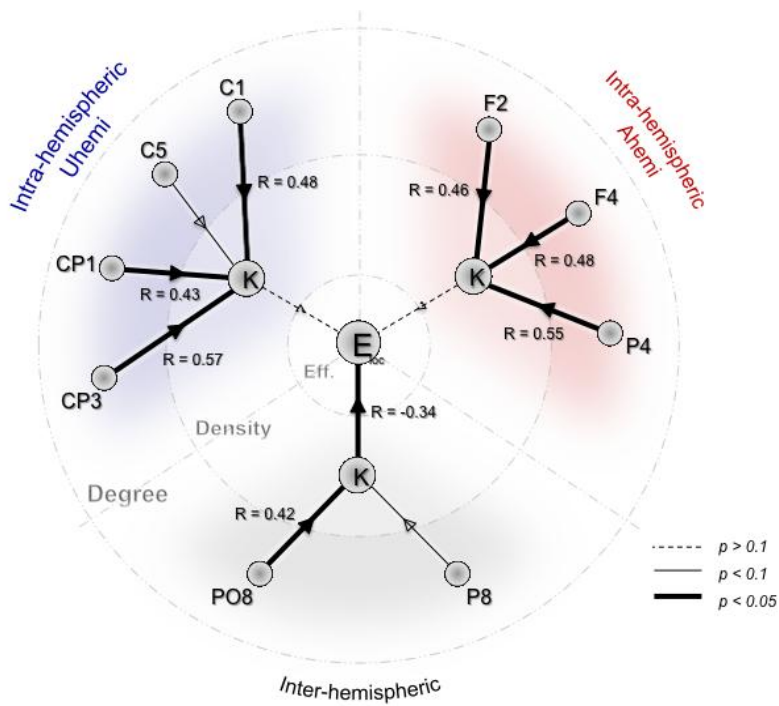

**Figure 5 - Graphical representation of statistical dependences between the changes of brain network properties at different topological scales in the Beta band.**

*Arrows indicate the network indexes used as independent (source node) and dependent (target node) variable in the performed linear regression analysis. Bold lines indicate significant dependences between network indexes. The thicker is the line the higher is the significance. The regression coefficients R are also reported as the weight of each link.*

*Nodes are spatially organized in three different concentric layers according to the topological scale they belong to. Small-scale indexes are arranged in the outer layer (intra-degree, $D_{intra}$; upper, left region of the outer layer for the unlesioned -Uhemi- hemisphere and upper, right region of the same layer for the lesioned -Ahemi- hemisphere), intermediate-scale indexes (intra-degree, $K_{intra}$) are in the middle layer, and large-scale indexes (local efficiency, $E_{loc}$) are in the inner layer. Inter-hemispheric indexes at the intermediate and small-scales (i.e. inter-density $K_{inter}$ and inter-degree $D_{inter}$) are illustrated in the lower region of the middle layer.*

As a consequence, being the small-world index *SW* the product of $E_{glo}$ and $E_{loc}$ (see section **2.3.1**), we concluded that the decrease in *SW* (green line) was due primarily to the fall in $E_{loc}$. Eventually, by superimposing the average $K_{inter}$ values for the actual brain networks in the *Beta* band, we determined that the *MI* of the affected hand was characterized by greater interhemispheric connectivity and lower local efficiency compared with the unaffected condition.

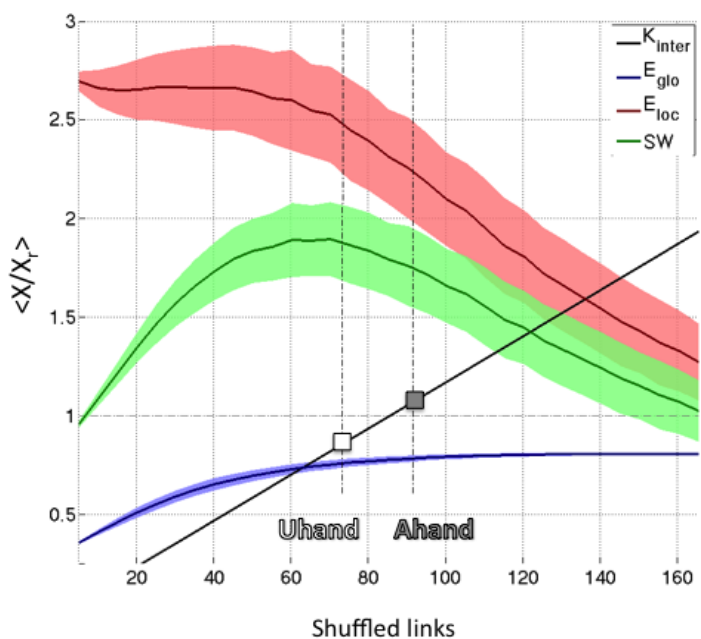

**Figure 6 - Intrinsic relationships between interhemispheric density and efficiency-based indexes.**
*The different colored lines show the mean profiles of the $K_{inter}$ (black) $E_{glo}$ (blue), $E_{loc}$ (red), and SW (green) values, as generated by the simulation model with the parmeter $p_{inter}=0.8$ (see section 2.4.1). The respective standard deviations are represented by the same-colored patches. All the computed vaues are divided by the mean of 1000 equivalent random graphs (Y-axis) as previously done for the actual brain network values (section 2.3.4). On the x-axis the increasing number of shuffled links is displayed.*

*The mean values of $K_{inter}$ estimated in the Beta band for the actual brain networks are superimposed on the simulated values. The white square stands for the unaffected hand condition (MI Uhand), while the grey square represents the affected condition (MI Ahand).*

3.3 Correlation with functional/clinical motor scales

A robust statistical correlation was observed between the connectivity of the affected hemisphere in the *Beta* band and the level of functional motor impairment of the stroke patient sample, as measured using the Fugl-Meyer Assessment (FMA) scale. Specifically, the FMA correlated positively ($R=0.53, p=0.045$) with the variation in $K_{intra}$ in *Ahemi* between conditions, based on the difference $\Delta K_{intra}(Ahemi)$ (**Fig. 3E**)—thus, higher FMA values reflected greater $K_{intra}(Ahemi)$ values in Ahand versus Uhand. $\Delta K_{intra}(Ahemi)$ also correlated positively with the European Stroke Scale (ESS) and Medical Research Council (MRC) scale scores, albeit insignificantly (**Fig. S2**). No other significant correlations were reported.

# 4 Discussion

In this study, we demonstrated that *EEG*-derived brain networks experience specific topological modifications during *MI* of the affected hand, with respect to the unaffected, occurring primarily within the *Beta* frequency oscillation (*13-30 Hz*). *Beta* rhythms during *MI* have been correlated to sensorimotor processes in healthy subjects (McFarland et al., 2000) and stroke patients (Gerloff et al., 2006). In the latter, the relative asymmetry between ipsilesional and contralesional hemispheres during contralateral *MI* and its relationship with functional motor impairment remain debated (Graziadio et al., 2012; Shahid et al., 2010). Here, the involvement of beta oscillatory activity is consistent with its function in the processing of overt and covert motor performance and corroborates previous observations of changes in *EEG* activity after stroke.

**4.1 Reduced smallworldness**

The recruitment of the affected hand *MI* was characterized by a significant reduction in smallworldness, *SW,* compared with the unaffected hand, reflecting a critical deviation from what might be considered optimal brain organization (Gerloff and Hallett, 2010). This decline was concomitant with a decrease in local efficiency $E_{loc}$ (**Fig. 3B**), meaning that there was a lower tendency of the *EEG* network to form tightly knit groups. In the context of networks, the presence of highly connected nonoverlapping groups implies that the network divides naturally into sets of nodes with dense connections internally and sparser connections between sets.

Based on the anatomic substrate of the human brain network, comprising 2 predetermined macro groups of regions— ie, here the affected and unaffected hemispheres—one might wonder to what extent the reduction in local efficiency could be predicted, based on the increase in interdensity $K_{inter}$ between hemispheres (**Fig. 3C**) or the decrease in intradensity $K_{intra}$ within contralateral hemispheres (**Fig. 3D**). Our regression analysis indicated that the decrease in local efficiency depended significantly on the rise in interhemispheric density, whereas there was no relationship between the changes in intrahemispheric densities (**Fig. 5**). This dependence was corroborated by the simulation model, which demonstrated that the reduction in local efficiency was caused mechanically by the increase in interhemispheric links (**Fig. 6**). Thus we hypothesize that the loss of smallworldness (ie, large-scale attribute) during *Ahand* was determined primarily by the abnormal rise in connectivity between hemispheres (ie, intermediate-scale attribute).

**4.2 Enhanced interhemispheric connectivity**

A recent *EEG* study demonstrated a significant flow of information from the contralesional to ipsilesional sensorimotor cortex in the lower *Beta* frequency band during movement of a paralyzed hand in less recovered stroke patients (Serrien et al., 2004) .

In our study, we observed a significant increase in interdensity $K_{inter}$ during *MI* of the affected hand (see **Fig. 3C**), which depended significantly ($p<0.01$) on the rise in interdegree $D_{inter}$ of site *PO8* in the ipsilesional hemisphere (**Fig. 4A**). This result implicates the contribution of local connectivity between the parietooccipital region in the affected hemisphere (*PO8* matched) and contralesional hemisphere. Whether this outcome reflects a functional reorganization that involves the "intact" hemisphere directly to compensate for the affected sensorimotor cortex is unknown.

Previous studies, based on *EEG* and *MEG* recordings, have suggested that *Beta* synchronous networks that involved the posterior parietal lobe constitute a general mechanism that implements attentional preparation, or readiness (Gross et al., 2004). Thus, the higher interhemispheric connectivity in the parieto-occipital region of the ipsilesional hemisphere could result instead from greater attentional resource engagement for patients during covert motor performance that involves their affected upper limbs. Notably, increased attention on the task has been postulated to represent a compensatory mechanism in cases of incomplete recovery (Strens et al., 2004).

**4.3 Abnormal recruitment of the affected hemisphere**

One of the most common issues after unilateral stroke is how it alters the function of the ipsilesional hemisphere (Grefkes and Fink, 2011; Westlake and Nagarajan, 2011). Compared with the unaffected hemisphere, in which the contralateral *MI* effects significantly greater intrahemispheric connectivity with respect to the ipsilateral condition, the affected hemisphere experiences a similar, but weaker, difference between intrahemispheric connectivity that was elicited by the contralateral and ipsilateral *MI*. This result suggests that some form of hemispheric lateralization was preserved in our sample of subacute stroke patients (1-3 months).

However, whereas contralateral intrahemispheric connectivity depended significantly on the local connectivity ($D_{intra}$) of the typically involved primary sensorimotor areas in *Uhand* (McFarland et al., 2000) (**Fig. 4B,** left hemisphere), the contralateral intrahemispheric density depended significantly on the node degrees of the frontal (*F2* and *F4* matched) and parietal (*P4* matched) areas in *Ahand* (**Fig. 4B,** right hemisphere). Collectively, these findings implicated abnormal recruitment of the cortical regions of *Ahemi* during the respective contralateral *MI*.

A previous study, showed that when the ipsilesional primary motor cortex is no longer capable of making functional contributions, secondary regions gain importance and perhaps become necessary components to sustain further, albeit incomplete, recovery (Calautti et al., 2001). In particular, the frontal and parietal regions are involved in implementing attentional function during preparation in cue-based tasks (Fischer et al., 2010). However, whether the influence of these regions is compensatory due to enhanced attention to the task or marks a true reorganization of the motor cortical network is unknown.

**4.4 Correlation with motor impairment**

For the affected hemisphere (*Ahemi*), the imbalance in $K_{intra}$ between contralateral (*Ahand*) and ipsilateral (*Uhand*) *MI* was significantly related to the severity of the functional impairment in the upper limb (FMA)—less extensive impairments effected a greater imbalance in favor of the contralateral *MI* condition (see **Fig. 3E**). Such a mechanism could reflect partial preservation of the original hemispheric laterality, similar to what was observed in the unaffected hemisphere. Yet, this laterality was less evident in patients with more severe motor impairments, resulting in lower sensibility of the affected hemisphere in recruiting a dense *FC* in the affected hemisphere in *Ahand.*

This finding points out a potential *EEG* network-based biomarker for the assessment of cortical plasticity that is related to the degree of motor impairment after unilateral stroke. $\Delta K_{intra}(Ahemi)$ could be used, for instance, to assess functional reorganization of the brain during specific *MI*-based rehabilitation strategies, such as those that exploit brain-computer interface (*BCI*) systems, which provide the patient with an online measure of his *MI* activity and devises appropriate neurofeedback to establish a more ecological and closed-loop neuromotor rehabilitation (Buch et al., 2012; Pichiorri et al., 2011).

**4.5 Methodological considerations**

*4.5.1 Task-based connectivity*

Baseline resting states are fundamental in understanding most complex systems (Raichle et al., 2001). The advantages and disadvantages of task-related connectivity with respect to resting-state connectivity in stroke were discussed recently (Westlake and Nagarajan, 2011). Essentially, task-based paradigms elicit brain activity that do not reflect the underlying anatomic structure exclusively, but they can generate many confounding effects that are difficult to isolate (Damoiseaux et al., 2006). In this study, we deliberately chose a specific task paradigm, consisting of the *MI* of hand grasping. *MI* has been the focus of many neuroscience groups for its significance in revealing the neural substrate of cognitive components of movement (Lotze and Halsband, 2006). Recently, a large debate was developed on the possible function of *MI* in sustaining motor recovery after stroke, renewing interest on this subject (Ietswaart et al., 2011).

*4.5.2 Different lesioned hemispheres*

One shortcoming of this study is that 9 of *20* patients suffered from lesions to the right hemisphere, whereas the remaining 11 had a left hemispheric stroke. Thus, differences in lesion location might have introduced confounding variables with regard to the common outcome interpretation. To limit this variability, we created a combined group of patients where only their right hemisphere could be considered affected. Although we are aware of the drawbacks of this approach, we report that it is performed frequently to avoid oversegmentation of the study population and create a group of patients that is sufficiently large to provide statistically valid results and conclusions (Luft et al., 2004; Ward et al., 2004). However, in a separate statistical analysis, there were no significant differences in the *Beta* band between the left (*n=11*) and right hemisphere-lesioned subgroups (*n=9*) in terms of network indices under the *Ahand* or *Uhand* condition (**Tab. S2**).

*4.5.3 Unaffected side as control condition*

Another issue is related to our experimental design in which we considered the *MI* task performed with the non paralyzed hand as a reference condition (*Uhand*) with respect to our target condition, namely the *MI* of the affected hand (*Ahand*). Previous studies have shown the existence of alterations of the unaffected hemisphere motor cortical output after stroke (Shimizu et al., 2002) that however, had little or no influence on functional motor recovery (Netz et al., 1997). In our experimental condition, we assumed that possible pathological changes of the non-affected hemisphere did not generate any functional motor impairments of the contralateral hand as indicated by the functional motor assessment. Bearing in mind that our *Uhand* condition cannot be a substitute of a normal control condition (ie, motor imagery performed by healthy age-matched individuals), the robust and sensible differences observed in the brain networks associated with the 2 experimental conditions performed by the same patient are valuable to allow insights into the picture of the brain responses to stroke lesions and their functional relevance.

*4.5.4 Subcortical and cortical lesions*

To exclude the fact that the obtained results could be somehow influenced by possible *EEG* signal modifications due to presence of anatomical lesions under the scalp electrodes, we repeated the analysis for the subgroup of subcortical patients (n=5, see Table S1). Even if the small sample size prevented a proper statistical computation, the results obtained for the subcortical group showed a tendency which is similar to what observed in the overall group analysis (see Fig. **S3** and **S4**). More importantly, the remaining subgroup of subcortical and/or cortical patients (n=15) still exhibited a pattern of significant differences between conditions that is comparable to the results obtained from the entire group of 20 patients (**Fig. S5** and **S6**).

# 5 Conclusions

This report is the first advanced network analysis to describe multiscale topological attributes of *EEG* motor networks in the presence of a unilateral stroke lesion. Compared with the unaffected hand imagination, the *MI* of the affected hand was characterized by a brain network that displayed lower smallworldness and local efficiency (large-scale attributes) topology, increased interhemispheric connectivity (intermediate-scale attributes), and higher regional connectivity over the frontoparietal areas of the affected hemisphere in the *Beta* band. Using a simple, but novel, approach we identified the influence of the smaller scale on larger-scale network properties and represented them as a "*network of network properties.*"

The relevance of our findings seems to be confirmed by the observed correlation between the level of patients' motor impairments and the connectivity "status" of the affected hemisphere, thus suggesting the presence of a complementary neuromarker that could be monitored throughout longitudinal recording sessions and considered in future motor cognitive-based rehabilitation strategies.

# 6 Appendices

**A Network analysis**

The mathematical representation of a network is a graph. A graph consists of a set of nodes *N* and a set of links that weigh a certain relationship between vertices. The adjacency matrix *A* contains information on the connectivity structure of the graph and has dimensions *NxN*. When a link connects 2 nodes *i* and *j*, the corresponding entry of the adjacency matrix is $a_{ij} \neq 0$; otherwise, $a_{ij} = 0$. In our graphs, nodes represent scalp electrodes (*N=61*), and unweighted links represent a significant coherence value between pairs of *EEG* signals.

*Large-scale indices.* Global efficiency and local efficiency are based on the concept of efficiency $e_{ij}=1/d_{ij}$, which is the reciprocal of the distance $d_{ij}$, computed as the length of the shortest path that connects 2 vertices *i* and *j* (Latora and Marchiori, 2001). A small-world network is characterized by global and local efficiencies that are both relatively higher than random graphs with the same number of nodes and links. Such a configuration is considered optimal in terms of a "perfect" balance between the integration and segregation properties of brain function and efficient communication between brain regions (Tononi et al., 1994). Moreover, initial clinical studies in functional neuroimaging have demonstrated that different conditions or pathologies can alter the optimal small-world architecture of the functional brain network (Achard et al., 2012; De Vico Fallani et al., 2007; Liu et al., 2008; Stam et al., 2007).

*Intermediate-scale indices.* Given the adjacency matrix *A*, the interdensity $K_{inter}$ reads:

$$K_{inter} = \frac{1}{N_S^2} \sum_{i,j \in S_{Ahemi}, Uhemi} A(i,j) \qquad (A.1)$$

where $N_S^2$ is the total number of crossing edges between 2 sets of same cardinality $\#(N_{SAhemi}) = \#(N_{SUhemi}) = N_S$.

The intradenisty $K_{intra}$ of a set S reads

$$K_{intra}(S) = \frac{2}{N_S^2 - N_S} \sum_{i \neq j \in S} A(i,j) \qquad (A.2)$$

where $S = S_{Ahemi}, S_{Uhemi}$. By definition, the intradensity ranges from *0* to *1*. The higher that $K_{intra}(S)$ is, the more connected the nodes in group *S* are.

*Small-scale indices.* Given the adjacency matrix *A*, the interdegree $D_{inter}$ of a node *i* is:

$$D_{inter}(i) = \sum_{j \in S_{tot}} A(i,j) \delta(i,j) \qquad (A.3)$$

where $S_{tot} = S_{Ahemi} \cup S_{Uhemi}$; $i \in S_{tot}$; and $\delta(i,j) = 1$ if *i* and *j* belong to different groups, and $\delta(i,j) = 0$ when they belong to the same group.

The intradegree $D_{intra}$ of a node *i* is:

$$D_{intra}(i) = \sum_{j \in S} A(i,j) \qquad (A.4)$$

where $S = S_{Ahemi}, S_{Uhemi}$; and $i \in S$.

## B Network simulation model

Let $G(N,L)$ be an unweighted and undirected network with $N$ nodes and $L$ links. Divide $G$ into 2 subsets, $S_A$ and $S_B$, consisting of $N_A$ and $N_B$ nodes, respectively, where $N_A + N_B \leq N$. The objective of the simulation model is to generate a sequence of $L$ network configurations, in which each configuration has a number of links $l = 1, 2, \ldots L$ that are redistributed randomly within the network. Notably, the probability $p_{inter}$ that a shuffled link is reassigned between $S_A$ and $S_B$ is given by total number of connections between the subsets and the total number of connections of the entire network, whereas the probability $p_{intra}$ that the same link is reassigned within 1 of the 2 subsets is given by total number of connections within the subsets and the total number of connections of the entire network.

Without loss of generality, assume that $N$ is even and $N_A = N_B = N/2$. In this simple case, it is easy to prove that $p_{inter}$ reads:

$$p_{inter} = \frac{N}{2(N-1)} \qquad (B.1)$$

and the probability $p_{intra}$ reads:

$$p_{intra} = \frac{N-2}{2(N-1)} = 1 - p_{inter} \qquad (B.2)$$

In our model, we wanted to parameterize $p_{inter}$ to modulate the amount of shuffled links to be redistributed within and between $S_A$ and $S_B$. Thus, our parameterized probability is defined as:

$$p_{inter} = \frac{N}{2(N-1)} + c, c \in \left( \frac{-N}{2(N-1)}, \frac{N-2}{2(N-1)} \right) \qquad (B.3)$$

It follows from Eq. (B.2) and Eq. (B.3) that when the parameter $c = -N/2(N-1)$, then $p_{inter} = 0$ and $p_{intra} = 1$, meaning that all $l$ links are reassigned randomly only within the 2 subsets; when the parameter $c = (N-2)/2(N-1)$, then $p_{inter} = 1$ and $p_{intra} = 0$, meaning that all $l$ links are rearranged exclusively between the 2 subsets.

## Acknowledgments

This work is partially supported by the European ICT Programme Project FP7-224631(TOBI). We would like to thank Claudia Di Lanzo for her preliminary computations, Giovanni Tessitore and Roberto Prevete for their supplementary analyses, and Mario Chavez for his useful suggestions on the modeling part. This paper only reflects the authors' views; funding agencies are not liable for any use that may be made of the information contained herein.